\documentstyle[aps,prd]{revtex}

\input{psfig}
\begin{document}
\twocolumn

\title{Behavior of the diffractive cross section in
hadron-nucleus collisions}

\author{{M. Batista, 
R. J. M. Covolan  and A. N. Pontes} \\ 
{\small Instituto de F\'{\i}sica {\em Gleb Wataghin}}\\
{\small Universidade Estadual de Campinas, Unicamp}\\
{\small 13093-970\ Campinas \ SP \ Brazil} \vspace{0.4cm}\\
\begin{minipage}{15cm}
A phenomenological analysis of diffractive dissociation of nuclei in 
proton-nucleus and meson-nucleus collisions is presented. The theoretical 
approach employed here is able to take into account at once data of the 
HELIOS and EHS/NA22 collaborations that exhibit quite different atomic mass 
dependences. Possible extensions of this approach to hard diffraction in 
nuclear processes are also discussed.
\vspace{0.1cm} 
\vspace{0.1cm}\\
To be published in Phys. Rev. C
\end{minipage}}
\maketitle
\section{Introduction}

Nowadays, a lot of attention is being dedicated to experimental and
theoretical studies of hadron diffractive interactions with the 
primary goal of investigating the Pomeron
and its role in both soft and hard  processes. 

However, dissociation reactions in which nuclei are the diffractively 
excited 
objects are not so well known. In fact, only two experiments have dealt
with this kind of process until now. The first one \cite{helios} is the
analysis of the diffractive dissociation of Be, Al, and W in collisions 
with 450 GeV/c protons, performed by the HELIOS Collaboration at the 
CERN Super Proton Synchrotron. The second one \cite{ehsna22} was performed 
at the same accelerator by the EHS/NA22 Collaboration, which has analyzed 
the diffractive dissociation of Au and Al excited by a 250 GeV/c meson 
beam composed of $\pi^+$ and $K^+$.

Probably the most striking result coming out of these experiments is that
the measured target-diffraction cross sections $\sigma_{TD}$ revelead to
have quite 
different atomic mass ($A$) dependences. Expressing 
this dependence as $\sigma_{TD} \propto A^\alpha$, 
the HELIOS Collaboration has obtained $\alpha = 0.35 \pm 0.02$ (that is, a 
value close to 1/3), whereas the EHS/NA22 Collaboration has found 
$\alpha = 0.58 \pm 0.06$ (not far from 2/3, which corresponds to the typical 
$A$ dependence in nuclear inelastic interactions).

In a recent paper \cite{batista1}, two of us have presented an analysis 
of the proton inclusive spectrum obtained in proton-nucleus collisions 
based on a multiple scattering formalism.  In such an analysis, particular 
attention was payed to the diffractive component of the inelastic cross 
section $p A$ resulting in a quite good description of the experimental
data of the HELIOS Collaboration.
In the present paper, we extend the previous analysis to the diffractive 
dissociation of nuclei in meson-nucleus collisions. 

\section{Theoretical framework}

Our starting point is 
the discussion of the diffractive dissociation cross section for 
hadron-proton ($hp$) collisions (in this case, $h = p,\ \pi^+,\ K^+$).
In the diffractive region, the invariant cross section of an inclusive
process of the type  $h p \rightarrow h' X$ is given by
\begin{equation}
E\frac{d^3\sigma}{d{\bf p}^3} =
\frac{s}{\pi} \frac{d^2\sigma}{dt\ dM^{2}_{X}},
\label{eq2a}
\end{equation}
with
\begin{equation}
 -t = m_{h'}^2\ {(1-x)^2}/{x} + {p_T^2}/{x}, 
\end{equation}
where $x = 2 p_L /{\sqrt s}$ is the Feynman variable of the particle 
$h'$ and 
$M^{2}_{X}$ is the invariant mass diffractively excited, which is 
defined as 
$M^{2}_{X} \equiv (p_h + p_p - p_{h'})^2$. Usually variable 
$\xi = M^{2}_{X}/s = 1-x$ is also employed to describe this sort of 
process.

In the approach used here, the single diffractive cross section,
Eq.~(\ref{eq2a}), is expressed in terms of a modified version of 
the Triple Pomeron model \cite{collins} which was adequate to
prevent against unitarity violation (see \cite{goulianos}):
\begin{equation}
\frac{d^{2}\sigma_{SD}}{d\xi\ dt} (h p \rightarrow h X) = 
f_{R}^h (\xi,t) \times \sigma_{{\tt I\!P} p}(s\ \xi) 
\label{Eq.1}
\end{equation}
where $f_{R}^h$ is the {\it renormalized} Pomeron flux factor 
\cite{goulianos}, that is 
\begin{equation}
f_{R}^h(\xi,t)=\frac{f_{S}^h (\xi,t)}{N(s)} 
\label{Eq.2}
\end{equation}
with the {\it standard} flux factor given by the Donnachie-Landshoff 
expression \cite{donnachie}\begin{equation}
f_{S}^h (\xi,t)=\frac{\beta^{2}_{h}}{16\pi}\ F^{2}(t)\ 
\xi^{[1-2\alpha_{{\tt I\!P}}(t)]}
\label{Eq.3}
\end{equation}
and
\begin{equation}
N(s)=\int_{1.5/s}^{1}\int_{-\infty}^{0}f_{S}^h (\xi,t)\ dt\ d\xi. 
\label{Eq.4}
\end{equation}
In Eq.~(\ref{Eq.3}), $\beta_{h}$ stands for the hadron-pomeron couplings
at the quasi-elastic vertices, whose values can be obtained directly from
the analysis of the total cross sections reported in
Ref.~\cite{covolan}. With a suitable change of normalization these 
values result 
to be $\beta_{p}=6.82$ $GeV^{-1}$, $\beta_{\pi}=4.13 \ GeV^{-1}$ and 
$\beta_{K}=3.71$ $GeV^{-1}$. In the same equation, $F(t)$ corresponds to the 
hadron form factor. When a proton is the recoiling particle, $F(t)$ is 
given by the Dirac form factor, 
\begin{equation}
F_1(t) = \frac{(4m^2_p - 2.79t)}{(4m^2_p - t)}\ 
\frac{1}{(1-\frac{t}{0.71})^2}; 
\end{equation}
for pions and kaons, we apply the dipole formula 
\begin{equation}
F_{1}(t)=\frac{1}{(1-\frac{t}{\mu^{2}_{m}})^{2}}, 
\label{Eq.5}
\end{equation}
with $\mu^{2}_{\pi}=0.92$ $GeV^{2}$ and $\mu^{2}_K=1.10$ $GeV^{2}$.

In Eq.~(\ref{Eq.1}), the pomeron-proton cross section is written
as  
\begin{equation}
\sigma_{{\tt I\!P}p}(M^{2}_{X})=\beta_{p}\ g_{{\tt I\!P}}\ (s\ \xi)^{\epsilon},
\label{Eq.6}
\end{equation}
where $g_{{\tt I\!P}}=0.87$ $GeV^{-1}$. The pomeron trajectory is always 
$\alpha(t)=1+\epsilon+\alpha't$, with $\epsilon=0.104$ and $\alpha'=0.25$ 
$GeV^{-2}$ obtained in \cite{covolan}. Eqs.~(\ref{eq2a})-(\ref{Eq.6}) 
completly specify how one calculates the diffractive cross section 
for the processes $p p \rightarrow p X$, $\pi^+ p \rightarrow \pi^+ X$
and $K^+ p \rightarrow K^+ X$.

\section{Nuclear diffractive dissociation}

Now, we turn our attention to diffraction in nuclear collisions.		
As proposed in \cite{frichter}, the invariant cross section for the 
inclusive reaction $h A \rightarrow h X$ is expressed in terms of 
\begin{equation}
\frac{d^{3}\sigma}{dx\ dp^{2}_{T}}(hA{\rightarrow}hX) = 
\sum_{\nu=1}^{A}\ \sigma^{hA}_{\nu}\ D^{N}_{\nu}(x,p^{2}_{T}),
\label{Eq.7}
\end{equation}
where $\sigma^{hA}_{\nu}$ is the partial inelastic cross section 
resulting of $\nu$ interactions: 
\begin{equation}
\sigma^{hA}_{\nu}=\int d^{2}b\ \frac{A!}{\nu!(A-\nu)!}\ P^{\nu}_{A}(b)\ 
[1-P_{A}(b)]^{A-\nu}.
\label{Eq.8}
\end{equation}
$P_{A}(b)$ is the probability of a nucleon to suffer an inelastic 
interaction at a given impact parameter {\em b}, which is expressed in 
terms of the nuclear density through the relationship   
\begin{equation}
P_{A}(b)=\sigma_{inel}^{hp}\int_{-\infty}^{+\infty} dz\ \rho_{A}(z,b).
\label{prob}
\end{equation}

The nuclear densities applied here are the same as those 
used in \cite{batista1} and respect the normalization condition
$\int d^3r\ \rho_A (r) = 1$. For light nuclei ($6 \leq A \leq 18)$ they 
follow the expression\footnote{We notice that
a factor $1/A$ is missing in the parametrization corresponding to 
Eq.~(\ref{rho1}) in the Appendix of Ref.~\cite{batista1}.} 
\begin{equation}
\rho(r) = \frac{4}{\pi^{3/2}\ {a_0^3}\ A} 
\left[1+\frac{1}{6}(A-4)\frac{r^2}{a_0^2} \right] \exp{(-r^2/{a_0^2})},
\label{rho1}
\end{equation}
with $a_0=[(r_0^2-r_p^2)^{1/2}/(5/2-4/A)]^{1/2}$, $r_0=1.2\ A^{1/3}$ fm, 
and $r_p=0.8$ fm. 
For heavier nuclei ($A\geq 18$), $\rho(r)$ is calculated according 
to the Woods-Saxon formula \cite{barret}, that is  
\begin{equation}
\rho({r})=\frac{c_0}{1 + exp[{(r - r_0)}/{b_0}]},
\label{rho2}
\end{equation}
where $c_0$ is the normalization constant  
\begin{equation}
c_0=\frac{3}{4\pi r_0^3}\ \frac{1}{1+(b_0 \pi/r_0)^2}
\end{equation}
and $b_0=0.4$ fm.

In order to calculate the inelastic cross sections $\sigma_{inel}^{hp}$ 
appearing in Eq.~(\ref{prob}), we introduce the following parametrizations
(already in milibarns) which are based on Regge phenomenology:

\begin{equation}
\sigma^{p p}_{inel} = 12.37\ s^{0.104} + 34.90\ s^{-0.20} - 31.30\ s^{-0.54},
\label{sigin}
\end{equation}
\begin{equation}
\sigma_{inel}^{\pi p} = 7.86\ s^{0.104} + 20.49\ s^{-0.20} - 8.02\ s^{-0.54},
\label{Eq.13}
\end{equation} 
\begin{equation}
\sigma_{inel}^{K p} = 6.46\ s^{0.104} + 20.46\ s^{-0.20} - 27.3\ s^{-0.54}.
\label{Eq.14}
\end{equation} 

The distribution $D^{N}_{\nu}(x,p^{2}_{T})$ is generated by a recurrence 
formula that starts with the assumption (see \cite{batista1} for details)
\begin{eqnarray}
D_{\nu=1}^{N}(x,p^{2}_{T})=
\frac{1}{\sigma^{hp}_{inel}}\ 
\left(\frac{d^3\sigma}{dx\ dp^{2}_{T}}\right )^{hN\rightarrow pX}.
\label{dsigpp}
\end{eqnarray}
To obtain the {\it total} inclusive cross section for the process
$h A \rightarrow h X$, the sum over $\nu$ in Eq.~(\ref{Eq.7}), which 
represents the number of times that the incident particle is scattered 
in the nuclear environment, should in principle run until its maximum 
value\footnote{In practice, not all terms need to be calculated in order 
to obtain a quite good description of the data (see \cite{batista1}).}.

However, when only diffractive events are concerned, it is enough to put 
$\nu=1$ since these processes are supposed to take place through 
single peripheral interactions with the outlying nucleons. 
Based on this argument, we use Eqs.~(\ref{Eq.7}) and (\ref{dsigpp}) 
to express the single diffractive component of the invariant 
cross section in hadron-nucleus collisions as 
\begin{equation}
\frac{d^{2}\sigma_{SD}}{d\xi\ dt}(h A{\rightarrow}h X) = 
\frac{\sigma^{hA}_{\nu=1}}{\sigma^{hp}_{inel}}\ 
\frac{d^{2}\sigma_{SD}}{d\xi\ dt}(hp{\rightarrow}hX),
\label{Eq.15}
\end{equation}  
which can be entirely calculated from what has been established previously.
Integrating Eq.~(\ref{Eq.15}) over $\xi$ and $t$, one obtains 
for the single diffractive hadron-nucleus cross section a very simple 
relationship:
\begin{equation}
\sigma^{hA{\rightarrow}hX}_{SD} = 
\frac{\sigma^{hA}_{\nu=1}}{\sigma_{inel}^{hp}}\ 
{\sigma^{hp{\rightarrow}hX}_{SD}}.
\label{Eq.16}
\end{equation}

\section{Results and Discussion}

A comparison of the model outlined above with cross section data 
has already been presented in Ref.~\cite{batista1} for the case of 
$p A \rightarrow p X$ with a slightly different parametrization. 
Here we show in Fig.~1 the differential cross 
section as a function of the invariant
mass $M_{X}^{2}$ compared to the EHS/NA22 data \cite{ehsna22} which 
are given in terms of the normalized distribution ${1}/{N_{ev}}\left 
({dN_{ev}}/{dM^{2}_{X}}\right )$ for the reactions 
$(\pi^+/K^+)\ Al \rightarrow (\pi^+/K^+)\ X$ and 
$(\pi^+/K^+)\ Au \rightarrow (\pi^+/K^+)\ X$ (in these data there is
no distinction between pions and kaons). 
\vspace{0.25cm}
\begin{figure}[htbp]
\centerline{\psfig{figure=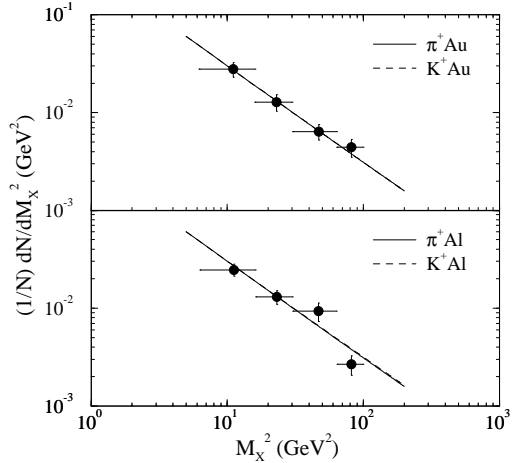,height=5.0cm}}
\vspace{0.65cm}
\caption{{Normalized $M^2_X$ distributions. 
The experimental data are from {\protect{\cite{ehsna22}}}}}
\end{figure}

In Fig.~2, theoretical predictions for the cross section 
$\sigma^{hA{\rightarrow}hX}_{SD}$, given by Eq.~(\ref{Eq.16}) with 
$h = p, \pi^+, K^+$, are compared 
with data of both collaborations, HELIOS \cite{helios} and EHS/NA22 
\cite{ehsna22}. As can be seen, the model provides a quite good description
for the $A$ dependence of the cross section for the various reactions.

This last result can be better understood if we construct a toy model
by assuming for the nuclear densities the (over)simplified form,
\begin{equation}
\rho(r) = \frac{1}{\pi^{3/2}\ d_0^3}\ e^{-r^2/d_0^2}.
\end{equation}
In such a case, all calculations can be performed analitically and
the nuclear diffractive cross section results to be
\begin{equation}
\sigma^{hA{\rightarrow}hX}_{SD} = 
\pi\ d_0^2\  \left[1 - (1 - \frac{\sigma_{inel}^{hp}}{\pi\ d_0^2})^A \right]\ 
\frac{\sigma^{hp{\rightarrow}hX}_{SD}}{\sigma_{inel}^{hp}}.
\label{toy}
\end{equation} 

Although such a formula is not appropriate for quantitative analysis, 
it is quite interesting from a qualitative point of view because it 
gives us a hint of how the ``elementary" inelastic cross section 
distinguishes the $A$ dependence of the different reactions: 
the higher $\sigma_{inel}^{hp}$ is, the flatter the $A$ dependence of the term 
within square brackets becomes. Based on this argument, the behavior 
observed in Fig.~2 is easily understood since $\sigma_{inel}^{pp} > 
\sigma_{inel}^{\pi^+ p} > \sigma_{inel}^{K^+ p}$. Of course, the real
calculation is not as simple as Eq.~(\ref{toy}), however the influence of 
$\sigma_{inel}^{hp}$ in the $A$ dependence of the term $\sigma^{hA}_{\nu=1}$ 
in Eq.~(\ref{Eq.16}) is quite the same.

As we have seen, the phenomenological model presented here provides 
a satisfactory description for {\it soft} nuclear diffraction. Now, let 
us make some conjectures about how it might employed to make predictions 
on {\it hard} diffractive processes generated by nuclear collisions. 
Before going to the point, let us make a brief digression.

The possibility for diffractive interactions occur in {\it hard} regime 
was first proposed by the Ingelman-Schlein (IS) model \cite{ingsc}. 
According to this model, a diffractive reaction may take place as a two-step 
process in which (1) a Pomeron is emitted from the quasi-elastic vertex 
and then (2) partons of the Pomeron interact with partons of the 
other hadron giving rise to dijets production or any other hard process. 
In fact, the very concept of Pomeron flux factor, which appears in 
Eq.~(\ref{Eq.1}), was introduced to represent the first step of such a process, 
being defined as 
\begin{equation}
f (\xi,t) \equiv \frac{1}{\sigma_{{\tt I\!P} p}(s\ \xi)}\ 
\frac{d^{2}\sigma_{SD}}{d\xi\ dt}.   
\label{flux}
\end{equation}
The second step is calculated through the QCD parton model and requires
the knowledge of the Pomeron structure function. 
\vspace{0.3cm}
\begin{figure}[htbp]
\centerline{\psfig{figure=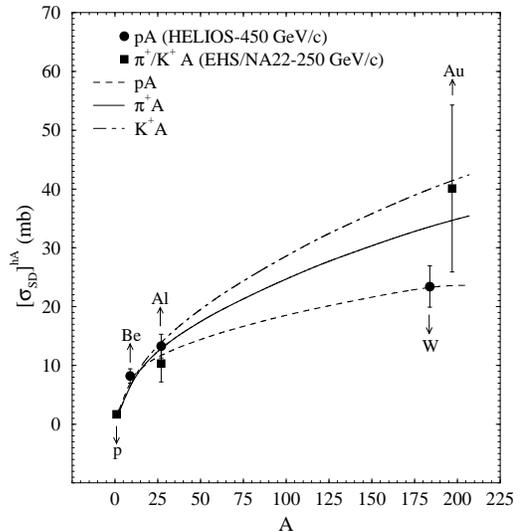,height=6.0cm}}
\vspace{0.55cm}
\caption{\label{g9} {Total cross section for nuclear diffractive dissociation. 
The elements corresponding to the nuclear targets are indicated by the arrows.
The data are from {\protect{\cite{helios}}} and {\protect{\cite{ehsna22}}}}}
\end{figure}

Discussion on these themes can be found, for instance, in two recent works 
which reported a study on the Pomeron structure function \cite{covolan2} 
and an extensive analysis of the role of the standard and renormalized 
Pomeron flux factors in the diffractive production of $W$'s and dijets 
at the DESY HERA and Fermilab Tevatron colliders \cite{covolan3}. 
This analysis \cite{covolan3} has shown that the 
standard flux factor is ruled out by the data and that the renormalized 
flux, which presents good results for diffractive hadroproduction, has 
 serious difficulties in diffractive photoproduction. In order
to overcome such difficulties, the IS model was reproposed in 
\cite{covolan3} in a modified form such that the Pomeron flux factor 
is replaced by a new distribution called {\it diffraction factor} which
is defined as 
\begin{equation}
F_{SD}(\xi, t) \equiv \frac{1}{\sigma_{SD}^{hp}}\ 
\frac{d^{2}\sigma_{SD}^{hp}}{d\xi\ dt}.
\label{diffac} 
\end{equation}
This new approach presents good results for both hadroproduction and 
photoproduction of dijets.

Turning to nuclear diffraction, we see that replacing Eq.(\ref{Eq.1}) 
into Eq.(\ref{Eq.15}) one gets 
\begin{equation}
\frac{d^{2}\sigma_{SD}}{d\xi\ dt}(h A{\rightarrow}h X) = f(\xi,t)
\times \frac{\sigma^{hA}_{\nu=1}}{\sigma^{hp}_{inel}} 
\ \sigma_{{\tt I\!P} p}(s\ \xi), 
\label{nucl}
\end{equation}
where the flux factor $f(\xi,t)$ was left free to assume whichever form
(standard or renormalized).

Now, considering how to apply the previous ideas to estimate the 
rates of some particular {\it hard} nuclear diffractive process (supposed 
to occur), one might think about two possibilities:

1) Sticking to the original IS model (namely, normalization of the  
differential cross section by $\sigma_{{\tt I\!P} p}$), one could 
define from Eq.(\ref{nucl}) a new flux factor that would depend on $A$, 
\begin{equation}
f_A (\xi,t) \equiv \frac{1}{\sigma_{{\tt I\!P} p}(s\ \xi)}\ 
\frac{d^{2}\sigma_{SD}}{d\xi\ dt}(h A{\rightarrow}h X) = 
f(\xi,t)\ \frac{\sigma^{hA}_{\nu=1}}{\sigma^{hp}_{inel}}; 
\label{fluxA1}
\end{equation}
2) On the other hand, it seems natural to define a sort of Pomeron-nucleus 
cross section by 
\begin{equation}
\sigma_{{\tt I\!P} A}(s\ \xi) \equiv 
 \frac{\sigma^{hA}_{\nu=1}}{\sigma^{hp}_{inel}} 
\ \sigma_{{\tt I\!P} p}(s\ \xi),
\label{sigA}
\end{equation}
and consider the flux factor defined by, 
\begin{equation}
f (\xi,t) \equiv \frac{1}{\sigma_{{\tt I\!P} A}(s\ \xi)}\ 
\frac{d^{2}\sigma_{SD}}{d\xi\ dt}(h A{\rightarrow}h X), 
\label{fluxA2}
\end{equation}
as the correct quantity to apply in the calculations. Of course, these
definitions would imply in quite different predictions.

Alternatively one could consider, instead of the flux factor, the idea of 
diffraction factor, Eq.(\ref{diffac}). Since the integral of this 
distribution is normalized to the unity by definition \cite{covolan3}, 
we see that its application to the hadron-nucleus case results to be the 
same as for the hadron-proton case, that is
\begin{equation}
F_{SD}(\xi, t)  = \frac{1}{\sigma_{SD}^{hA}}\ 
\frac{d^{2}\sigma_{SD}^{hA}}{d\xi\ dt} = \frac{1}{\sigma_{SD}^{hp}}\ 
\frac{d^{2}\sigma_{SD}^{hp}}{d\xi\ dt}
\label{diffacA} 
\end{equation}
once Eqs.~(\ref{Eq.15}) and (\ref{Eq.16}) are taken into account. Therefore,
with the concept of diffraction factor, there is no ambiguity about which 
expression to use.

It is important to notice however that, whatever be the choice, 
Eq.~(\ref{fluxA1}), (\ref{fluxA2}), or 
(\ref{diffacA}), the previous discussion establish the main elements needed 
to perform the calculations and make predictions of hard diffraction in 
nuclear collisions in the same spirit that has been done in {\cite{covolan3} 
for proton-antiproton interactions. The only new input necessary are 
parametrizations of structure functions obtained for nuclear environments.
In this sense what is being proposed above is a sort of extension of the 
Ingelman-Schlein model (in its original and modified forms) to take into
account hard diffraction in nuclear processes.

In summary, we have presented in this paper a phenomenological approach that
provides a quite good description of the apparently discrepant $A$ 
dependence observed in the diffractive cross sections of meson-nucleus and 
proton-nucleus collisions. The model used for such allows one to imagine 
possible theoretical frameworks within which one could make predictions
on hard diffractive processes originated from hadron-nucleus interactions.

\section*{Acknowledgments}

We would like to thank the Brazilian governmental agencies CNPq, CAPES 
and FAPESP for  financial support.


\begin{thebibliography}{9999}
\bibitem{helios} HELIOS Collab., T. {\AA}kesson {\it et al.}, 
Z. Phys. C {\bf 49}, 355 (1991).

\bibitem{ehsna22} EHS/NA22 Collab., N. M. Agababyan {\it et al.}, 
Z. Phys. C {\bf 72}, 65 (1996).

\bibitem{batista1} M. Batista and R. J. M. Covolan, Phys. Rev. C {\bf 60}, 
014902 (1999).

\bibitem{collins} P. D. B. Collins, {\it An Introduction to 
Regge theory
and high energy physics} (Cambridge University Press, 
Cambridge, England, 1977); P. D. B. Collins and A. D. Martin, 
{\it Hadron Interactions} (Adam Hilger Ltd., Bristol, England, 1984).

\bibitem{goulianos} K. Goulianos, Phys. Lett. B {\bf 358}, 379 (1995);
B {\bf 363}, 268 (E) (1995); K. Goulianos and J. Montanha, Phys. Rev. D 
{\bf 59}, 114017 (1999).

\bibitem{donnachie} A. Donnachie and P. V. Landshoff, Nucl. Phys. B 
{\bf 303}, 634 (1988).

\bibitem{covolan} R. J. M. Covolan, J. Montanha, K. Goulianos, 
Phys. Lett. {\bf B 389}, 176 (1996).

\bibitem{frichter} G. M. Frichter, T. K. Gaisser and T. Stanev, Phys. 
Rev. D {\bf 56}, 3135 (1997).

\bibitem{barret} R. C. Barret and D. F. Jackson, {\it Nuclear Sizes
and Structure} (Oxford University Press, Oxford, 1977).

\bibitem{ingsc} G. Ingelman and P. E. Schlein, Phys. Lett. B {\bf 152}, 
256 (1985).

\bibitem{covolan2} R. J. M. Covolan and M. S. Soares, Phys. Rev. D  
{\bf 57}, 180 (1998).

\bibitem{covolan3} R. J. M. Covolan and M. S. Soares, Phys. Rev. D  
{\bf 60}, 054005 (1999); Erratum: Phys. Rev. D {\bf 61}, 19901 (2000). 

\end{thebibliography}
\end{document}